\documentclass[pra,twocolumn,showpacs,preprintnumbers,
amsmath,amssymb,floatfix]{revtex4}
\usepackage{hyperref}
\usepackage{amsmath}
\usepackage[dvips]{epsfig}

\newcommand{\beq}{\begin{equation}}
\newcommand{\eeq}{\end{equation}} 
\newcommand{\beqa}{\begin{eqnarray}}
\newcommand{\eeqa}{\end{eqnarray}}
\newcommand{\ba}{\begin{array}}
\newcommand{\ea}{\end{array}}

\begin{document}
\title{Dipolar Bose gas in highly anharmonic traps}
\author{Francesco Ancilotto$^{1,2}$ and Flavio Toigo$^{1,2}$} 
\affiliation{$^1$Dipartimento di Fisica e Astronomia 
"Galileo Galilei" and CNISM, Universit\`a di Padova, 
Via Marzolo 8, 35122 Padova, Italy \\
$^2$CNR-IOM Democritos, via Bonomea, 265 - 34136 Trieste, Italy} 

\begin{abstract} 
By means of mean-field theory, we have studied the 
structure and excitation spectrum 
of a purely dipolar Bose gas in pancake-shaped trap where
the confinement in the x-y plane is provided by a highly
anharmonic potential resulting in an almost uniform confinement in
the plane. We show that the stable condensates is characterized  
by marked radially structured density profiles.
The stability diagram is calculated by independently varying the
strength of the interaction and the trap geometry. 
By computing the Bogoliubov excitation spectrum near the instability line
we show that soft "angular" rotons are responsible for the collapse
of the system.
The free expansion of the cloud after the trap is released 
is also studied by means of time-dependent calculations, showing that 
a prolate, cigar-shaped condensate is dynamically stabilized
during the expansion, which would otherwise collapse.
Dipolar condensates rotating with sufficiently high
angular velocity show the formation of multiply-quantized giant vortices,
while the condensates acquire a ring-shaped form.
\end{abstract} 
\date{\today}
\maketitle

\section{\bf INTRODUCTION} 

In recent years, ultracold gases of dipolar particles (which include
atoms with  large magnetic moments and polar molecules)
have attracted a great deal of interest because of their
peculiar properties, which are due to the presence of the
anisotropic, long-range dipole-dipole interaction in 
addition to the usual short-range correlation interaction.

A dipolar Bose Einstein Condensate (BEC) has been 
first realized with a gas of $^{52}$Cr
atoms \cite{sthuhler}, where the dipolar interaction 
energy between magnetic moments was about $15\%$
of the short-range interaction energy as 
calculated on the basis of the scattering length.
Dipolar systems made of polar molecules have been achieved 
as well\cite{weinstein}.
Since then, purely dipolar condensate have been realized by tuning
to zero the scattering length. The properties of dipolar BEC have been
the subject of numerous experimental and theoretical studies, extensively reviewed  
in Ref.\cite{lahaye} and Ref.\cite{baranov}.

The long-range nature of the dipolar interaction 
together with its anisotropic character leads in these system to the
appearance of a rich variety of phenomena, whose properties depend
crucially upon the shape of the trapping potential and of the
interaction strength
\cite{sthuhler,werner,santos,baranov,yi,odell,ronen,malo}.

A purely dipolar Bose gas (i.e. one where no short-range 
correlation interaction among the atoms is present,
or where the dipole-dipole interaction is much larger than 
the contact interaction) 
is always unstable in the spatially uniform case. 
This can be seen by recalling that in a completely 
polarized uniform condensate of density $n_0$, 
where all dipoles (each of magnitude $d$) are
parallel to each other, the dispersion law for 
elementary excitations as derived within the
Bogoliubov theory, is given by\cite{lahaye}:
\beq
\epsilon ({\bf k})=[ E_k^2 + 2 E_k n_0 C_{dd}(cos^2\theta _k-1/3) ]^{1/2}
\label{dispersion}
\eeq
where $E_k=\hbar ^2k^2/2M$ and $\theta _k$ is the angle between the
excitation momentum ${\bf k}$ and the common direction of the dipoles.
$C_{dd}$ is the dipole-dipole interaction strength, proportional
to $d^2$ (see the following).  
The instability is clearly seen from the fact that at small $k$
and $cos^2\theta _k<1/3$ the excitations energies become imaginary.

In a completely polarized condensate confined by a cigar-shaped (prolate) trap 
elongated along the direction of dipoles
alignment, the interaction
will be mainly attractive and the condensate is unstable 
towards collapse,
similarly to the case of a cold gas with attractive short-range
interactions (negative s-wave scattering length).
Conversely, in pancake-shaped (oblate) traps  the 
tendency of the dipoles to align along
the polarization axis in order to increase their 
attractive head-to-tail interaction is counteracted by the 
strong confinement in this direction, which favors instead
configurations with atoms or molecules laying in a plane orthogonal   
to their dipole moments. This effect may make the 
dipolar interaction energy predominantly repulsive, 
leading to stabilization of the BEC when the trap is oblate enough.

Therefore the dipolar gas offers the possibility 
of modifying the effective 
atom-atom interaction by modifying the trap geometry,
which can easily be controlled in  experiments.

It has been shown\cite{santos1} that pancake dipolar condensates
may exhibit a roton-maxon feature in the excitation spectrum. 
The presence, position and depth of the roton minimum
can be tuned by varying the density,
the confining potential and the short-range interaction strength.
Since the superfluid critical velocity, as obtained from the 
Landau's argument, is reduced in the presence 
of a roton minimum,
this opens the possibility of manipulating the superfluid
properties of trapped condensates.

When the roton gap disappears the condensate becomes 
intrinsically unstable:
since the instability occurs at a specific value of the momentum, 
this points to the possibility of realizing 
a non-uniform ground-state. The associated "self-assembled" 
density modulations,
that are expected in the high density regime,
can have a supersolid character\cite{josserand}.
In spite of this exciting possibility,  
no conclusive evidence has been gathered
so far \cite{komineas} that a 
supersolid phase of dipolar BEC 
actually exists.

A purely dipolar (i.e. the s-wave scattering length is $a=0$) 
condensate in pancake-shaped traps
is stable\cite{koch}, provided
the number of atoms/molecules is not too large.
The stability diagram for dipolar condensates in
pancake-like traps has been calculated in Ref.\cite{ronen},
by using the time-dependent Gross-Pitaevskii (GP) equation.

In a narrow region of the stability diagram, bi-concave shaped
("blood cell")
condensate wavefunctions have been found\cite{ronen,dutta}, 
with the maximum 
density away from the center of the cloud.
Such effect is due to the long-range, mainly repulsive
forces in oblate traps, in a way very much similar to the 
accumulation of charges on the surface of conducting materials.
These peculiar structures are predicted to exist
in a very narrow portion of the stability diagram,
and also very close to the instability line.
The soft-mode which is responsible for the collapse 
of these bi-concave condensates when the instability 
edge is reached is characterized by azimuthal density oscillations
with angular momentum quantum number $m>0$ ("angular" rotons), 
as opposed to the
soft modes which drive the collapse of 
a "normal" condensate with isotropic interactions (where the maximum
density is at the center), where $m=0$ instead. 

The observation of such bi-concave states is experimentally difficult
because they have a small contrast (i.e. small density inhomogeneities),
and moreover the region in the parameter space where they exist covers a very small area, very close
to the instability line.
The structure and dynamics of bi-concave dipolar condensates
has been theoretically investigated using the GP equation 
near the threshold for instability, 
and a possible experimental signature of their appearance
have been proposed
in Ref.\cite{wilson1}, in the form of non trivial 
angular distribution of the products of the 
system once the collapse is induced by varying
the scattering length.

The superfluid character of dipolar condensate is signalled by
the formation of quantized vortices in a rotating 
condensate. For this reason, the theoretical study of vortex
states in dipolar systems has been pursued by several 
groups.
In general, rotation of a harmonically trapped dipolar gas is
found to affect the stability of the 
condensate\cite{wilson,yi2,eberlein,abad,dell,dell1,malet}.
Vortex states in dipolar systems in toroidal traps have been studied 
as well\cite{abad1}, where an azimuthal dependence of a particle
density was found, characterized by symmetry-breaking density profiles
where the density is inhomogeneously distributed along the torus.

The collapse dynamics of a $^{52}$Cr
dipolar condensate, induced by a sudden reduction of the
s-wave scattering length 
characterizing the contact interaction
below a threshold value, has been 
studied in Ref.\cite{metz,parker}. 
The collapse dynamics, triggered either by
an adiabatic or nonadiabatic change in the 
ratio between the dipolar and the contact
interaction strength, has been theoretically 
studied in Ref.\cite{parker},
where both "global" (i.e. characterized by highly 
elongated or flattened shapes of the atomic cloud) 
and "local" (i.e. characterized 
by inhomogeneous density profiles like shells, disks or
stripes) regimes of collapse were observed.

A rather complex spatial 
pattern characterizes the dynamics of the cloud expansion 
and collapse reported in Ref.\cite{metz},
involving an anisotropic, d-wave symmetric explosion of a
spherical condensate.
The collapse of pancake-shaped clouds 
has been studied in Ref.\cite{metz1}, showing similar
qualitative behavior.
The collapse dynamics as imaged during the 
experiment is quantitatively reproduced by 
numerical simulations based on the GP equation
without any adjustable parameter\cite{metz}.
The collapse 
of disc-shaped Bose dipolar gases has also been 
numerically studied in real-time dynamics in Ref.\cite{adhik}.

At variance with all the above investigations, 
where the confinement has always been chosen 
to be harmonic, even if anisotropic,
in the following we will study a purely 
dipolar BEC confined by 
pancake-shaped trap potentials characterized by a 
tightly harmonic confinement along the z-direction
(corresponding to the dipole polarization axis) 
but highly anharmonic in the x-y plane ("flat" confinement).
This choice is suggested by the fact that in such a trap, 
easily implemented experimentally, the system motion 
in the x-y plane is almost unaffected by the trap details 
within the region of confinement, while the residual effect
is due to the steep confinement at the system boundaries. 
The combination of these effects enhances the 
characteristic features associated with the non-local dipolar
interaction with respect to the case of harmonic confinement.
In fact we will show that, as a result of such confinement, 
highly structured densities of the dipolar BEC are obtained,
generalizing the "blood cell" condensate shapes 
predicted for harmonic confinement.
These highly structured radial density profiles 
might be easily observed experimentally
since they are present in a wide portion 
of the stability diagram.

We use the GP equation to compute both static and dynamical properties 
of these systems. The validity of mean-field approximation 
was tested using many-body Monte Carlo methods\cite{ronen1,blume} 
and found to 
provide a correct description of the dipolar gas in the dilute 
limit $na^3\ll 1$, i.e. at very low densities and/or away 
from shape resonances. 

We will determine the stability diagrams 
of the dipolar condensate as both
the strength parameter and the trap aspect ratio 
are varied, and show that the systems undergoes collapse 
for sufficiently large coupling parameter and/or sufficiently
large aspect ratios.

We find that the lowest excitation mode which becomes 
soft close to the instability has an azimuthal dependence proportional 
to $sin(3\phi)$ and $sin(4\phi)$: 
the condensate is thus expected to collapse
with density modulations in the angular coordinates
which break the cylindrical symmetry characterizing the
ground-state structures.

We have also studied the behavior of the condensate 
when subject to a rotation with a constant angular velocity
$\Omega $ around an axisparallel to the polarization direction:
for sufficiently high values of $\Omega $ the dipolar
condensate develops multiple quantized vortices.
Eventually, a single giant, multiply
quantized vortex appears, while the condensate density 
acquires the shape of a narrow torus.

Finally, we have studied the free 
expansion of the "blood-cell" condensate
after the confining trap is released and find
that the expansion is characterized
by an elongation into a cigar-shaped
dipolar condensate that would otherwise be unstable
towards collapse under stationary conditions.
Such dynamical stabilization of a cigar-shaped,
purely dipolar condensate should be easily imaged 
in experiments. 

\section{\bf METHODS AND CALCULATIONS} 

We assume that all the $N$ atoms (molecules) of the system
are in a Bose-Einstein condensate
described by the wavefunction $\Phi ({\bf r})$.
The direction of dipole polarization is assumed to be
along the $z$-axis.
The energy of the system at the mean-field level
is expressed by the functional
\begin{eqnarray}
E=\int {\Big [}{\hbar ^2 \over 2M} {|\nabla 
\Phi ({\bf r})|^2}+V_{t}({\bf r})|\Phi ({\bf r})|^2
+{g\over 2}|\Phi ({\bf r})|^4{\Big ]}d{\bf r}+
\nonumber \\
+{1\over 2}\int {\int {V_{dd}(|{\bf r}-
{\bf r}^\prime |)|\Phi ({\bf r})|^2  }
|\Phi ({\bf r}^\prime )|^2d{\bf r}\,d{\bf r}^\prime}
\end{eqnarray}
\label{energy}

The above functional includes a kinetic pressure term,
a term describing the trap potential used to confine the
system, the short-range ("contact") mean-field interaction energy,
and the dipole-dipole interaction potential:

\beq
V_{dd}({\bf R})={C_{dd}\over 4\pi} {(1-3cos^2\theta)\over R^3}
\eeq
where
${\bf R}={\bf r}-{\bf r}^\prime $ and $\theta $ is the
angle between the vector ${\bf R}$ and the polarization direction $z$.

In the case of polar molecules the prefactor is
$d^2/\epsilon _0$, where $d$ is the electric dipole moment
of the molecule and $\epsilon _0$ is the permittivity of the vacuum.
In the case of magnetic interaction the prefactor is instead
$\mu _0 \mu _d^2$, where $\mu _d$ is the magnetic dipole moment of each
atom and $\mu _0$ is the permeability of free space.

The minimization of the above energy functional
leads to the following Euler-Lagrange equation
\beq
\hat {H}\Phi ({\bf r}) = \mu \Phi({\bf r})
\label{stationary}
\eeq
where $\mu $ is a Lagrange multiplier whose value is determined
by the normalization condition $\int |\Phi ({\bf r})|^2 d{\bf r}=N$,
and
\begin{eqnarray}
\hat {H} \equiv 
-{\hbar ^2 \over 2M  }\nabla ^2 +V_{t}({\bf r}) +g|\Phi ({\bf r})|^2
+\nonumber \\
+{C_{dd}\over 4\pi} 
\int   
{(1-3cos^2\theta)\over (|{\bf r}-{\bf r}^\prime|)^3}
|\Phi ({\bf r}^\prime)|^2
d{\bf r}^\prime
\label{hamilt}
\end{eqnarray}

We will consider in the following a purely dipolar condensate, i.e. 
we take from now on $g=0$, implicitly assuming that the interparticle 
interaction is dominated by the dipole-dipole forces
($d^2>>|g|=4\pi \hbar ^2|a_s|/M$, where $a_s$ is the s-wave
scattering length).

We consider here the case where the trapping potential
is highly anharmonic ("flat") in the x-y plane, 
and harmonically confined along the polarization direction $z$:

\beq
V_{t}({\bf r})={M\over 2}\omega _z ^2z^2+\kappa (x^2+y^2)^{\gamma /2}
\label{trap}
\eeq
where $\gamma >2$.

Highly anharmonic potentials can be realized in principle
with the current laser optic techniques, as suggested
by the theoretical calculations of Ref.\cite{jaouadi}.
The resulting trapping potential behaves as
$V_l(\rho )\propto \rho ^{2l}$, where $\rho =\sqrt{x^2+y^2}$
is the radial distance from the beam center.
Very flat confining potentials in the x-y plane 
are obtained for large values of $l$,
although already for $l>4$ the condensate is characterized
by an almost constant density over the entire
trap volume in the case of purely repulsive BEC\cite{jaouadi}.
The possibility of studying a spatially uniform quantum gas
by loading the gas in a three-dimensional, quasi-uniform
potential has also been demonstrated in Ref.\cite{gaunt}.
The leading-order correction to the flatness of the box
potential was found to be $\propto \rho^{13\pm 2}$, 
which is equivalent to a flat potential 
for most many-body studies. 

In the following, we will take $\gamma =10$ 
in the expression (\ref{trap}) for the trap potential.

Upon scaling lengths by $a_z\equiv \sqrt{\hbar/M\omega _z}$ and 
energies by $\hbar \omega _z$, and
also by imposing that the dimensionless wavefunction
is normalized to unity, i.e. 
$\Phi ({\bf r})=\sqrt{N/a_z^3}\tilde {\Phi }(\tilde {{\bf r}})$
the equation (\ref{stationary}) can be written in dimensionless form as:

\begin{eqnarray}
{\Big [}-{{\tilde \nabla} ^2 \over 2  } + {1\over 2}\tilde{z}^2+
({a_z \over r_0})^{\gamma +2}{1 \over 2  }(\tilde{x}^2+\tilde{y}^2)^{\gamma /2}
{\Big ]}\tilde {\Phi }(\tilde {{\bf r}})
+\nonumber \\
+D \int {(1-3cos^2\theta) \over (|{\tilde {\bf r}}-{\tilde {\bf r}}^\prime |)^3} 
|{\tilde \Phi }({\tilde {\bf r}}^\prime )|^2
d{\tilde {\bf r}}^\prime 
\tilde {\Phi }(\tilde {{\bf r}}) = 
\tilde {\mu} \tilde {\Phi }(\tilde {{\bf r}})
\label{dimenseq}
\end{eqnarray}

Here $r_0\equiv (\hbar ^2 / 2M\kappa )^{1/(\gamma +2)}$
is the characteristic length for the  
anharmonic potential, expressed in terms of the
parameter $\kappa $ appearing in Eq.(\ref{trap}).
>From equation (\ref{dimenseq}) it appears that
two dimensionless parameters determine the 
solutions of the GP equation for a purely dipolar BEC
in the trapping potential (\ref{trap}):

\beq
\lambda ={a_z \over r_0} \,\,\,\,\,\,\,\,\,\,\,\,\,\,\,\,\,D=NC_{dd}M/4\pi \hbar ^2 a_z
\eeq

$\lambda $ controls
the aspect ratio of the trap while 
$D$ controls the dipole-dipole interaction strength.

This equation will be solved numerically,
to yield the lowest 
energy state $\Phi $ describing the condensate
in the ground-state for a given pair ${\lambda,D}$.
We do this by
propagating it in imaginary time, i.e. by solving the equation
\beq
{\partial \Phi \over \partial t}+(\hat {H}-\mu )\Phi ({\bf r}) = 0
\label{imag}
\eeq
The wave function $\Phi({\bf r})$
is represented on a three-dimensional uniform mesh in real space,
with periodic boundary conditions imposed on the system.
The cell must be wide enough to avoid spurious dipole-dipole 
interactions
between the atomic clouds and its periodically repeated
images\cite{ronenbis}.
The starting wavefunction is chosen in the form of a narrow 
gaussian placed in the center of the trap (although we verified that the final,
minimum energy structure does not depend upon a
particular choice for the initial wavefunction).

To compute the spatial derivatives
appearing in the GP equation (\ref{stationary}), 
we used an accurate 13-point finite-difference formula \cite{pi}.
The convolution integral in the potential energy term 
of Eq.(\ref{dimenseq}) is 
efficiently evaluated in reciprocal space 
by using Fast Fourier transform techniques,
and by recalling that the Fourier transform 
of $V_{dd}$ is\cite{lahaye}

\beq
{\tilde V}_{\bf k}={C_{dd} \over 3}(3cos^2\alpha -1)
\eeq
where $\alpha$ is the angle between ${\bf k}$ and
the $z$-axis.

In order to compute the excitation spectrum, we make 
the usual Bogoliubov transformation to a 
Hamiltonian describing a collection of non-interacting
quasi-particles for which the condensate is the vacuum:
\beq
\Psi ({\bf r},t)=e^{-i\mu t/\hbar}
[\Phi ({\bf r})+u({\bf r})e^{-i\omega t}
-v^\ast ({\bf r})e^{i\omega t} ]
\label{bogo}
\eeq
where $u({\bf r})$ and $v({\bf r})$ 
are the wavefunctions of the excitation mode
and $\Phi ({\bf r})$ 
is the solution of Eq.(\ref{stationary}).

The computation of the excited states 
is particularly demanding from a computational 
point of view if performed, as we did in the 
present work, in three-dimensional cartesian coordinates
(an alternative approach, exploiting the
cylindrical symmetry of the problem, has been developed
in Ref.\cite{ronenbis}).
It is thus important, in order to reduce the
computational burden, to use the smallest possible
number of points in the spatial mesh that guarantee
an accurate representation of the ground-state
wavefunction $\Phi $.
By studying the convergence in energy of the 
solution with increasing number of
points in the mesh, we verified that a relatively 
coarse grid with spacing $\Delta x \sim 0.3\,a_z$
is enough to accurately describe $\Phi({\bf r})$.

Because of our use of Fourier Transform techniques,
which imply that periodic boundary conditions 
must be imposed in our calculations,
we can expand the wavefunction $\Phi $ and
the complex functions $u,v$ in the form 
appropriate to a periodic system:
\begin{eqnarray}
\Phi ({\bf r})=\sum _{\bf G} \Phi_{\bf G} 
e^{i\bf G \cdot \bf r} \\
u({\bf r})=\sum _{\bf G} 
u_{ \bf G} e^{i\bf G \cdot \bf r} \\
v({\bf r})=\sum _{\bf G} 
v_{\bf G} e^{i\bf G \cdot \bf r}
\label{bloch}
\end{eqnarray}

In the above expansions, the ${\bf G}$-vectors are the
reciprocal lattice vectors appropriate to the 
space symmetry of the simulation cell containing the
dipolar cloud. In the present case, a simple
orthorhombic geometry is used, with a computational
cell of sides $L_x=L_y$ and $L_z$.
Substituting the expressions (\ref{bloch}) into the
time-dependent GP equation
$[i\hbar \partial /\partial t - \hat {H}]\Psi ({\bf r},t)=0$ 
associated with the Hamiltonian (\ref{hamilt}),
and keeping only terms
linear in the functions $u,v$, one obtains 
a set of equations for the coefficients
$\{u_{\bf G}\} $ and $\{v_{\bf G}\}$ that can be recasted into 
the following matrix form
\cite{anci}:

\begin{equation}\everymath{\displaystyle}
\begin{bmatrix}
{\bf A} & {\bf B}  \\
-{\bf B}  & -{\bf A}  \\
\end{bmatrix}\begin{pmatrix} {\bf u} \\ {\bf v}
\end{pmatrix}=\hbar \omega
\begin{pmatrix} {\bf u}
\\ {\bf v}   \,
\end{pmatrix}
\end{equation}
where the matrices
(with dimensions $(n_r^3\times n_r^3)$,
where $n_r$ is the real space mesh
used to integrate the stationary GP equation)
are defined as:

\begin{eqnarray}
{\bf A}_{{\bf G},{\bf G}^\prime}
\equiv \delta _{{\bf G},{\bf G}^\prime}[{\hbar ^2 {\bf G})^2
\over 2M}-\mu 
]\nonumber \\
+\tilde {U}_{\bf G-\bf G^ \prime}
+\sum _{\bf G^ {\prime \prime}}\Phi _{\bf G^ 
{\prime \prime}-\bf G^ \prime}
\Phi _{\bf G-\bf G^ {\prime \prime}}\tilde{V}_{
\bf G^ {\prime \prime}}
\label{matrix0}
\end{eqnarray}
\begin{eqnarray}
{\bf B}_{{\bf G},{\bf G}^\prime}
\equiv -\sum _{\bf G^ {\prime \prime}}
\Phi _{\bf G^ {\prime \prime}-\bf G^ \prime}
\Phi _{\bf G-\bf G^ {\prime \prime}}
\tilde{V}_{\bf G^ {\prime \prime}} 
\label{matrix}
\end{eqnarray}

The quantities $\tilde {U}_{\bf G}$ in Eqns.(\ref{matrix0}) 
are defined through 
\beq
V_{t}({\bf r})+\int{ V_{dd}(|{\bf r}-{\bf r} ^\prime|) |
\Phi ({\bf r} ^\prime)|^2d{\bf r} ^\prime }=
\sum _{\bf G} \tilde {U}_{\bf G}
e^{i{\bf G}\cdot {\bf r}}
\label{utilde}
\eeq

The excitation frequencies $\omega ({\bf k})$ 
can be determined from the solutions 
of the above non-Hermitian eigenvalue problem.
This can be reduced to a non-Hermitian problem 
of {\it half} the dimension 
(thus largely reducing the computational cost of diagonalization)
by means of a unitary transformation \cite{unitary}:
\beq
({\bf A}-{\bf B}) ({\bf A}+{\bf B}) |{\bf u}
+{\bf v}>=(\hbar \omega )^2
|{\bf u}+{\bf v}>
\label{unitary}
\eeq

If needed, one may calculate the separate ${\bf u}$, ${\bf v}$ 
by properly combining the
eigenvectors of Eq. (\ref{unitary}) with those 
of the associated eigenvalue problem
\beq
({\bf A}+{\bf B}) ({\bf A}-{\bf B}) |{\bf u}-{\bf v}>=(\hbar \omega )^2
|{\bf u}-{\bf v}> ,
\label{unitary1}
\eeq
again of reduced dimensions.

\section{\bf RESULTS AND DISCUSSION} 

We have solved the stationary GP equation in
imaginary time, as described in the previous Section, 
for different 
pairs of values $(\lambda,D)$: possible outcome of the 
calculations are either stable, converged states
or "collapsed" states where during the evolution
in imaginary time the density of the system catastrophically
shrinks within a narrow spatial region.

\begin{figure}
\epsfig{file=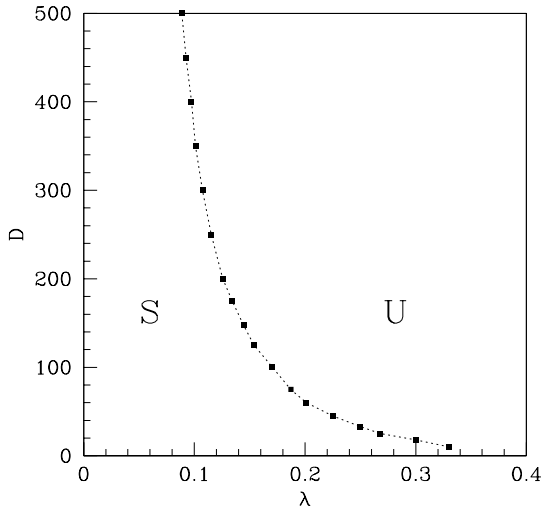,height=3.8 in,clip=}  
\caption{ Calculated stability diagram in the $D-\lambda $ plane.
}
\label{stability}
\end{figure}

The resulting stability diagram is shown in Fig.\ref{stability},
where a line in the $D-\lambda$ plane separates 
the stable (S) configurations
from the unstable (U) ones.
It appears that higher values of $D$ (strong dipole-dipole
coupling or large number of particles) and/or larger values of
the aspect ratio $\lambda $ (weaker confinement in the z-direction)
make the system unstable towards collapse.

\begin{figure}
\epsfig{file=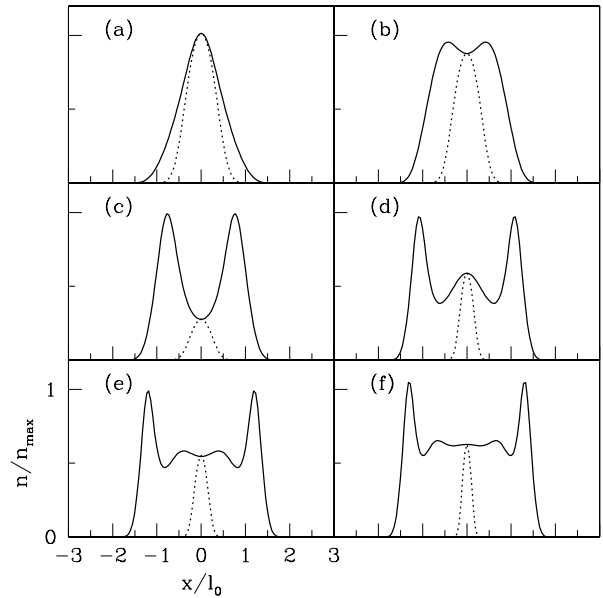,height=3.8 in,clip=}  
\caption{Density profiles (normalized to the maximum density
value) of stable 
configurations, shown along a cut in the x-y plane
across the trap center,
for different values of $D$ and $\lambda $.
(a):$D=10$, $\lambda =0.28$;
(b):$D=25$, $\lambda =0.24$;
(c):$D=50$, $\lambda =0.20$;
(d):$D=150$, $\lambda =0.13$;
(e):$D=250$, $\lambda =0.1$;
(f):$D=400$, $\lambda =0.08$.
The dotted lines show the density profile along the 
z-direction.}
\label{profiles}
\end{figure}

The density profiles associated with the
condensate structures in the stable part
of Fig.(\ref{stability}) are characterized by 
marked radial inhomogeneities,
and differ considerably from the usual 
appearance of harmonically trapped dipolar gas 
clouds, where the maximum
density is usually in the center of the trap.
We show in Fig.\ref{profiles} several structures obtained for
different pairs ($\lambda, D$). Here the number density $n$
is defined as 
$n({\bf r})=|\tilde {\Phi }({\bf r})|^2$.
Common to these structure 
is the accumulation towards the trap periphery, due to
the interaction between dipoles,
which results in 
a dense circular edge. Additional secondary peaks 
may appear for different values of ($\lambda, D$).
One may recognize in panel (c) a bi-concave ("blood cell")
structure like the ones discussed in the previous Section, 
but with a much higher contrast than those
realizable in harmonic traps.
Multiple radial peaks appear as well 
for different trap geometries and/or
strength parameter values.

\begin{figure}
\epsfig{file=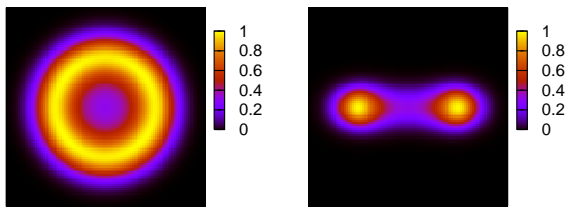,height=2.2 in,clip=}  
\caption{(color online) Equal-density map (shown in the x-y 
and x-z plane, respectively) 
for the condensate with D=50 and $\lambda =0.20$.
}
\label{dens_gs}
\end{figure}

In Fig.\ref{dens_gs} the ground state density profile of the
bi-concave condensate shown in panel (c) of Fig.(\ref{profiles}) 
is shown using equal density color maps
in the x-y and x-z planes, respectively.

\begin{figure}
\epsfig{file=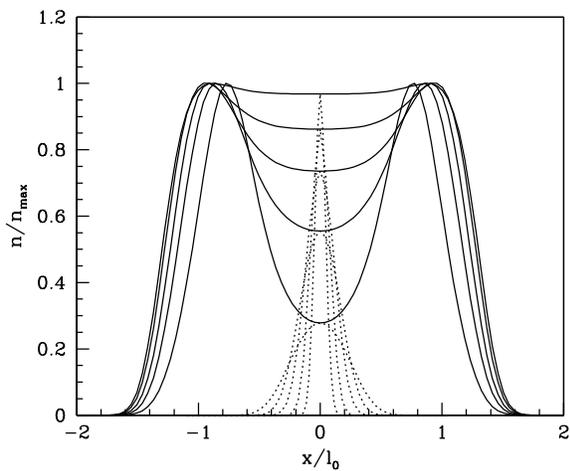,height=3.8 in,clip=}  
\caption{Density profiles (normalized to the maximum density
value) shown along the x-direction,
for $D=50$ and $\lambda =0.06,0.10,0.14,0.18,0.20$.
Approaching the instability line with higher values of $\lambda =$ correspond to more concave
structures.
The dotted lines show the density profile along the 
z-direction.}
\label{profiles_d_new}
\end{figure}

\begin{figure}
\epsfig{file=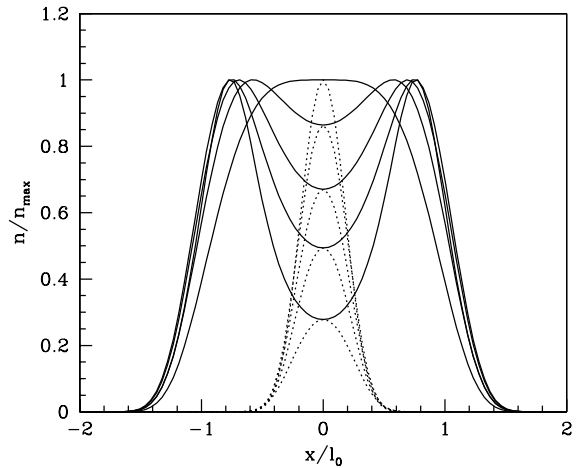,height=3.8 in,clip=}  
\caption{Density profiles (normalized to the maximum density
value) shown along the x-direction,
for $\lambda =0.20$ and $D=10,20,30,40,50$.
Approaching the instability line with higher values of $D$ correspond to more concave
structures.
The dotted lines show the density profile along the 
z-direction.}
\label{profiles_l_new}
\end{figure}

Unlike the case of harmonic confinement in the x-y
plane, where concave structures with maximum density
away from the trap center appears only in very small
portion of the stability diagram, in the present case
such shapes are easily realized with a variety of 
($\lambda $, $D$) pairs spanning a large portion of the stability region in
the diagram of Fig.(\ref{stability}). 
As an example, we show in Fig.(\ref{profiles_d_new}) and Fig.(\ref{profiles_l_new})
the ground-state density profiles 
for different choices of ($\lambda $, $D$).
Fig.(\ref{profiles_d_new}) shows the structures associated to a 
fixed value of $D$ and different values of
$\lambda $. It appears that lower values of $\lambda $
(i.e. corresponding to a tighter 
confinement of the dipolar gas along the z-direction)
result in less structured clouds.
Fig.(\ref{profiles_d_newbis}) shows instead the structures associated with a
fixed value of $\lambda $ and different values of the
interaction parameter
$D $. Less structured clouds now correspond to lower values
of $D$. 

Collective modes of 
dipolar condensates under harmonic confinement have been analyzed in
Ref.\cite{ronen} (see also Ref.\cite{baranov}).
As discussed in the Introduction, 
two possible solution of the stationary GP equation have been found:
a pancake shaped condensate (with the maximum of the
density at the trap center) or
a bi-concave shaped condensate which appears in a narrow region
of the stability region. In the pancake case, the mode
that drives the instability is a "radial" roton, i.e. 
it has a radial nodal pattern
($m=0$ projection of the angular momentum on the z-axis).
In a bi-concave condensate the mode going soft exhibits density modulations along the ring corresponding 
to non-zero $m$ ("angular" roton), 
with a subsequent breaking of the cylindrical symmetry\cite{ronen}.

For completeness, by using the Bogoliubov-deGennes equation formalism
discussed in the previous Section, we have also computed the excitation spectrum of selected
structures as a function of $D$ for a given value of $\lambda$, starting  below the stability line of Fig.(\ref{stability})
and increasing $D$, until the instability 
line is reached. 

The resulting excitation frequencies for the
case $\lambda =0.2 $ are shown in Fig.(\ref{dispersion}), where the 
calculated frequencies $\omega $ are shown as
a function of $D $. By approaching the instability value
it appears that several modes become soft. In particular, two
modes (those originating at $\omega =0.5\,\omega _z$ 
and $\omega =0.34\,\omega_z$ for $D=0$)
have frequencies that go to zero right at the edge of collapse.

The local density fluctuations 
associated with these excitation modes
can be calculated from

\begin{eqnarray}
\Delta n ({\bf r})=|u({\bf r})-
v({\bf r})|^2 \nonumber \\
\label{fluct}
\end{eqnarray}

\begin{figure}
\epsfig{file=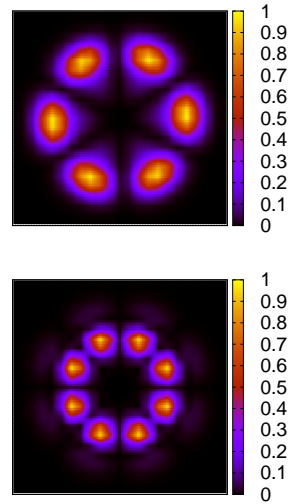,height=3.2 in,clip=}  
\caption{ (color online) Number density fluctuations associated with the
two soft modes shown in Fig.(\ref{dispersion}).
} 
\label{eigv}
\end{figure}

The density fluctuations $\Delta n ({\bf r}) $ 
(shown with a map of equal density colors in the x-y plane)
for the two soft modes which drive the collapse 
of the cloud are shown in Fig.(\ref{eigv}).

\begin{figure}
\epsfig{file=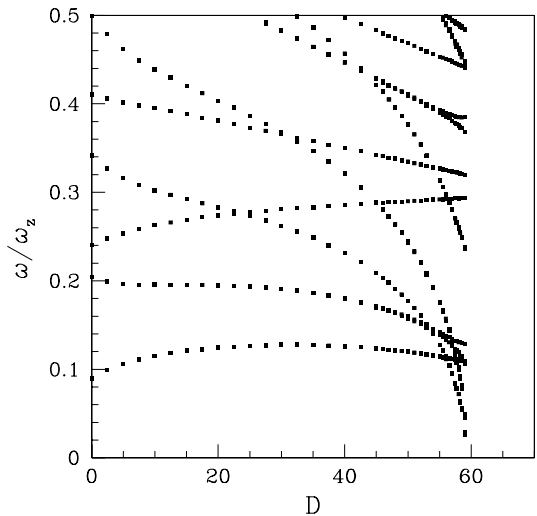,height=3.8 in,clip=}  
\caption{Excitation spectrum for dipolar condensates with $\lambda =0.2$,
as a function of the strength parameter $D$. 
}
\label{dispersion}
\end{figure}

>From Fig.(\ref{eigv}) it appears that
such modes have an azimuthal dependence proportional 
to $sin(3\phi)$ and $sin(4\phi)$: the 
condensate is thus expected to collapse
with density modulations in the angular coordinates
which breaks the cylindrical symmetry.
A similar breaking pattern has also been observed in the collapse
of harmonically trapped 
Bose clouds in pancake-shaped condensates from  
numerical simulations based on the GP equation\cite{adhik}.

We have also studied the character of the soft mode
responsible for the collapse of "normal" condensate 
clouds (i.e. with the maximum of the density at the center 
of the trap). One such configuration is shown in panel (a)
of Fig.(\ref{profiles}). In this case we find that the
mode that drives the instability  
has a radial nodal pattern
($m=0$ projection of the angular momentum on the z-axis),
similarly to what is found for harmonically trapped
condensates\cite{ronen,baranov}.

Next, we studied how the dipolar condensate evolves dynamically
after the trap is released.
The time evolution is obtained by solving the time-dependent 
GP equation:

\beq
i\hbar {\partial \Psi ({\bf r},t)\over \partial t}=
[\hat {H}-{i\hbar L_3\over 2}|\Psi |^4]\Psi  ({\bf r},t) 
\eeq

We have added to the dipolar GP hamiltonian
a dissipative term proportional to
$L_3\sim 2\times 10^{-40} $ m$^6$ s$^{-1}$
which 
describes three-body losses\cite{metz}.

\begin{figure}
\epsfig{file=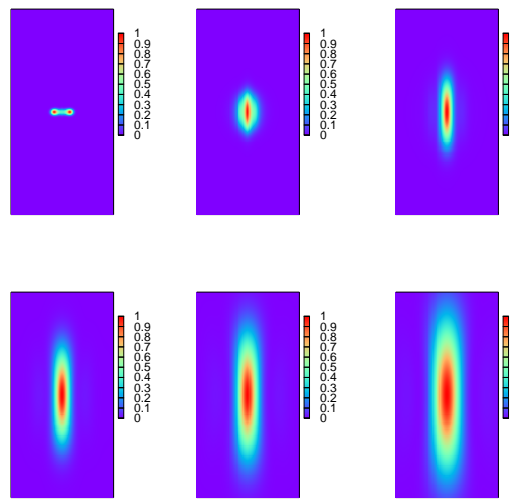,height=3.4 in,clip=}  
\caption{ (color online) Time evolution of the condensate after the release 
of the trap. From left to right and top to bottom,
the time instants are: 
$t=0,15,30,45,60,75,90\,\omega_z ^{-1}$
} 
\label{timev}
\end{figure}

\begin{figure}
\epsfig{file=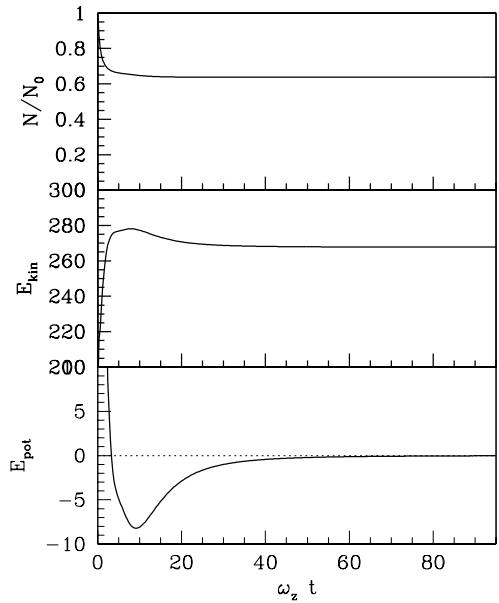,height=3.8 in,clip=}  
\caption{ Upper panel: number of atoms in the 
expanding condensate as a function of
time. Middle and lower panel: kinetic 
and potential energy as a function of time.
}
\label{expansion}
\end{figure}

We have used 
the Runge-Kutta-Gill fourth-order method \cite{rkg} 
to propagate in time the solutions of the previous equation.
The initial configuration is the ground-state structure
shown in Fig.(\ref{dens_gs}).
The time evolution after the trap is suddenly released is shown 
in the sequence of snapshots of Fig.(\ref{timev}).

The main feature in Fig.(\ref{timev}) is the radial inward
shrinkage of the initial state and the expansion along
the polarization axis, driven by the energy gain
associated by elongated configurations, and 
resulting in an inversion of the initial aspect ratio. 
In spite of the elongation towards 
a prolate geometry the condensate
is not collapsing because of the kinetic 
energy associated with the expansion. 
A similar inversion of the aspect 
ratio has been observed\cite{sthuhler} 
during the  
expansion of harmonically trapped gases when a repulsive
short-range interaction was present in addition to the dipole-dipole
interaction.
Here, however, we find a notable
dynamical stabilization of a cigar-shaped condensate
(at least during the expansion time)
occurring for a {\it purely dipolar}
BEC, which would be otherwise unstable.

In Fig.(\ref{expansion}) we show the number of atoms in the
condensate during the expansion (panel (a)).
The initial sudden loss is due to the sudden inward shrinkage
of the cloud in the radial direction. 
In panel (b) and (c) we show how the kinetic energy
and the dipole-dipole interaction energy vary
during the expansion. The potential energy, that is initially
positive due to the mainly repulsive character of the 
dipole-dipole interaction in the initially oblate condensate,
becomes rapidly negative, signalling the predominance 
of aligned dipole configurations as the shape changes 
from oblate to 
prolate, but then
smoothly decreases without sign of collapse.

We have verified, by comparing the above results 
with the expansion
of the same system but with the dipole-dipole interaction turned off, 
that the elongation is not simply a consequence of the 
tighter confinement along the z-direction and the associated
increased quantum pressure, but it is indeed energetically
driven by the dipole-dipole attraction: when such interaction
is not present, the expansion occurs in a more isotropic way.

To investigate the emergence of vortex structures in
our system, we search for the lowest-energy stationary states
solution of the time-dependent GP equation in a 
rotating frame-of-reference with constant 
angular velocity $\Omega $:

\beq
i\hbar {\partial \Psi ({\bf r},t)\over \partial t}=
[\hat {H}-\Omega L_z]\Psi  ({\bf r},t) 
\eeq
where $L_z$ is the z-component of the orbital angular momentum
operator.

\begin{figure}
\epsfig{file=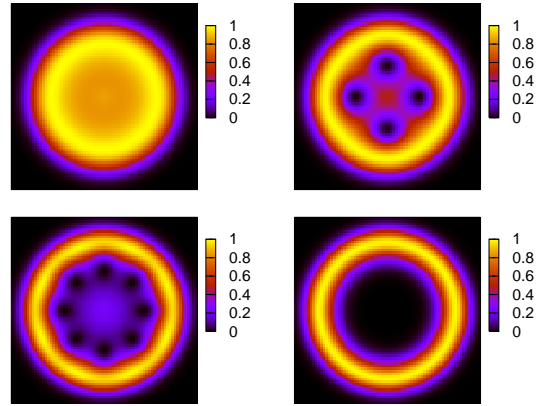,height=2.6 in,clip=}  
\caption{ (color online) Vortex structure for the condensate 
obtained with $D=50$ and $\lambda=0.11$,
for four different values of the 
rotational frequency $\Omega $.
>From left to right, and from top to bottom:
$\Omega =0.033,\,0.049,\,0.066,\,0.082\,\omega_z$
} 
\label{vort_all}
\end{figure}

We show in Fig.(\ref{vort_all}) our results for different values of $\Omega $.
Below a critical value $\Omega \sim 0.04\,\omega _z$ 
the condensate cloud is not affected
by the rotation. As soon as this critical angular velocity
is exceeded, multiple vortex structures develop in the inner portion
of the condensate where the density is lower, whose number
increases with the angular velocity. Eventually, they merge 
into a giant  
vortex which empties the inner region 
of the trap, leading to a ring-shaped
condensate.

Giant multi-quantized vortex states are known 
to be a feature of rotating BEC in the presence
of anharmonic trap potentials \cite{cetoli},
since anharmonic confinement improves
the stability of the system against 
centrifugal destabilization,
thus allowing to reach higher 
values of the angular velocity of
rotation than those allowed by harmonic confinement.

\section{\bf CONCLUSIONS} 

Within a mean-field approach, we have theoretically studied
static and dynamical properties of a purely dipolar, completely polarized
pancake-shaped Bose Einstein condensate subject to highly
anharmonic ("flat") confinement in the plane perpendicular to 
the dipole polarization axis. 

Marked radial density inhomogeneities   
are the distinguishing features of this system,
which generalize the "blood cell" cloud shapes observed  in
harmonically confined dipolar BECs \cite{ronen,dutta}.
We computed the stability diagram, showing that
such radially structured clouds appear in a 
wide variety of the parameters (dipole 
interaction strength and trap aspect ratio)
and thus should be experimentally observable.
The calculation of the excitation spectrum allows us to 
determine the type and symmetry of the soft
modes responsible for the collapse of the system,
which occurs in the form of "angular" roton excitations.

The free expansion of the dipolar gas 
is also studied, showing a temporary
stabilization, during the expansion process, of
a prolate cloud which would be unstable and thus
not be observable
under stationary conditions in a purely dipolar BEC.

Finally, the rotation of the trap is studied, and the 
ensuing formation of
multiple vortex structures. Eventually,
for sufficiently high angular velocities (which would
not be reachable in harmonic traps) a giant vortex 
develops with a wide empty core, and the BEC acquires
a ring-shaped geometry.

Since highly anharmonic traps can be realized with 
current state-of-the-art laser optical 
techniques, we expect that
most of the features described here will  
soon be confirmed by experiments. 

\medskip
\begin{acknowledgments}
We acknowledge useful discussions with 
Luca Salasnich and Maurizio Rossi.
\end{acknowledgments}

\end{document}